\documentclass{ws-procs10x7square}
\usepackage{amsfonts, amsmath, hyperref}
\newcommand{\upchi}{\raise1pt\hbox{$\chi$}}
\newcommand{\R}{{\mathord{\mathbb R}}}
\newcommand{\C}{{\mathord{\mathbb C}}}

\newcommand{\id}{{\mathord{\;\rm d}}}

\newcommand{\mfr}[2]{{\textstyle\frac{#1}{#2}}}

\newcommand{\an}{{\mathord{a}}^{\phantom{*}}}
\newcommand{\bn}{{\mathord{b}}^{\phantom{*}}}

\newcommand{\cA}{{\mathord{\mathcal A}}}

\newcommand{\cP}{{\mathord{\mathcal P}}} 
\newcommand{\cK}{{\mathord{\mathcal K}}}
\newcommand{\cH}{{\mathord{\mathcal H}}}
\newcommand{\cU}{{\mathord{\mathcal U}}}

\begin{document}

%%%%%%%%%%%%%%%%%%%%%%%%%%%%%%%%%%%%%%%%%%%%%%%%%%%%%%%%%%%%%% 
% title, author(s) and address(es) put here:                 %
%%%%%%%%%%%%%%%%%%%%%%%%%%%%%%%%%%%%%%%%%%%%%%%%%%%%%%%%%%%%%% 

\title{The energy of Charged Matter\thanks{Work partially supported by \uppercase{NSF}
    grant \uppercase{DMS}-0111298, by \uppercase{EU} grant HPRN-CT-2002-00277, by MaPhySto --
    A Network in Mathematical Physics and Stochastics, funded by The
    \uppercase{D}anish National Research Foundation, and by grants from the \uppercase{D}anish
    research council.  }}

\author{Jan Philip Solovej\thanks{On
    leave from \uppercase{D}epartment of \uppercase{M}athematics,
    \uppercase{U}niversity of \uppercase{C}openhagen,
    \uppercase{U}niversitetsparken 5, \uppercase{DK}-2100
    \uppercase{C}openhagen, \uppercase{D}ENMARK.  \newline \copyright 2003 by the
  author. This article may be
  reproduced in its entirety for non-commercial purposes.\hfill  }} \address{School of Mathematics\\
\normalsize Institute for Advanced Study\\
\normalsize 1 Einstein Drive \\
\normalsize Princeton, N.J. 08540\\
\normalsize {\it e-mail\/}: solovej@math.ku.dk }

\maketitle

\abstracts{In this talk I will discuss some of the techniques that
  have been developed over the past 35 years to estimate the energy of
  charged matter. These techniques have been used to solve stability
  of (fermionic) matter in different contexts, and to control the
  instability of charged bosonic matter. The final goal will be to
  indicate how these techniques with certain improvements can be used
  to prove Dyson's 1967 conjecture for the energy of a charged Bose
  gas---the sharp $N^{7/5}$ law.}

%%%%%%%%%%%%%%%%%%%%%%%%%%%%%%%%%%%%%%%%%%%%%%%%%%%%%%%%%%%%%
% The main text of your paper                                         %
%%%%%%%%%%%%%%%%%%%%%%%%%%%%%%%%%%%%%%%%%%%%%%%%%%%%%%%%%%%%%

\section{INTRODUCTION}

It is my aim in this contribution to review the main techniques
developed to study the problem of stability of matter or rather {\it
  stability} or {\it instability} of ordinary matter. By ordinary
matter I mean a macroscopic collection of charged particles (nuclei
and electrons) interacting solely through electrostatic or
electromagnetic forces.

The problem of stability of (ordinary) matter interacting through
electrostatic Coulomb forces was first formulated by Fisher and
Ruelle~\cite{FiRu} and proved in the seminal papers of Dyson and
Lenard~\cite{DyLe1,DyLe2}, although, of course the problem of stability of
individual atoms goes back to the origin of quantum mechanics. An
important assumption needed is that either the negatively or the
positively charged particles are fermions, i.e., obey the Pauli
exclusion principle. Without the fermionic assumption there is no
stability as proved by Dyson in \cite{Dy}.

The importance of the Pauli exclusion
principle for stability had been pointed out in the celebrated work of
Chandrasekhar \cite{Ch} on gravitational collapse and stability of
white dwarfs.
It was, however, not until the work of Dyson and Lenard that the
importance of the exclusion principle for the stability of ordinary
matter was emphasized.  In \cite{Dy} Dyson makes a very precise
conjecture regarding the nature of the instability without the
exclusion principle.

It is my ultimate goal here to discuss the proof of this
conjecture, which for the main part, is joint work with E.H.~Lieb.

Since the work of Dyson and Lenard there has been a lot of activity in
the area of stability of matter. Most celebrated is the work of Lieb and
Thirring~\cite{LiTh} giving an elegant proof with a bound of the
correct order of magnitude.  Several variations of the model have been
studied. Relativistic effects and magnetic interactions have been included.
I shall review some of these results below. In recent years the
attention has turned to studying the effect of quantizing the
electromagnetic field, what is often referred to as non-relativistic
quantum electrodynamics. I will not get into this recent development
here. 

\section{CHARGED MATTER IN QUANTUM MECHANICS}

The systems we discuss here are all given by $N$-particle Hamiltonians ($N$ being a positive integer) of the form 
\begin{equation}\label{eq:Ham}
H_N=\sum_{j=1}^NT_j+\sum_{1\leq i<j\leq
  N}\frac{z_iz_j}{|x_i-x_j|}+{\cU}.
\end{equation}
Here $z_j\in\R$ is the charge of particle $j$, $x_j\in\R^3$ is the
coordinate of particle $j$, and $T_j$ is the kinetic energy operator
for particle $j$. We shall here consider the kinetic operator to be of
one of the following four types.
\begin{enumerate}
\item {\bf The standard non-relativistic kinetic energy}
  $$
  T_j=-\frac{1}{2m_j}\Delta_j.
  $$
\item {\bf The relativistic kinetic energy}
  $$
  T_j^{\rm Rel}=\sqrt{-c^2\Delta_j+m_j^2c^4}-m_jc^2.
  $$
\item {\bf The magnetic kinetic energy}
  $$
  T_j^{\rm Mag}=
  \frac{1}{2m_j}\left(-i\nabla_j-\frac{z_j}{c}{\boldsymbol A}(x_j)\right)^2.
  $$
\item {\bf The magnetic Pauli kinetic energy}
  $$
  T_j^{\rm Pauli}=\frac{1}{2m_j}\left((-i\nabla_j-\frac{z_j}{c}{\boldsymbol
      A(x_j)})\cdot{\boldsymbol\sigma}_j\right)^2.
  $$
\end{enumerate}
Above, $m_j>0$ refers to the mass of particle $j$, ${\boldsymbol
  A}:\R^3\to\R^3$ is the vector potential for the magnetic field, and
${\boldsymbol\sigma}_j$ is the vector of $2\times2$ Pauli matrices for
particle $j$, i.e,  $ {\boldsymbol\sigma}=(\sigma_1,\sigma_2,\sigma_3) $, with 
$$
\sigma_1=\left( \begin {array}{lr} 0&\phantom{-}1\\\noalign{\medskip}1&0\end {array} \right),\quad
\sigma_2=\left( \begin {array}{lr} 0&-i\\\noalign{\medskip}i&0\end {array} \right),\quad
\sigma_3=\left( \begin {array}{lr} 1&0\\\noalign{\medskip}0&-1\end {array} \right).
$$
The Pauli kinetic energy acts on spinor valued functions (spin 1/2
particles), see the discussion of the relevant Hilbert spaces below.

The term $\cU$ in (\ref{eq:Ham}) represents the {\it magnetic field energy}
$$
\cU=\frac{1}{8\pi}\int_{\R^3}|\nabla\times{\boldsymbol
  A(x)}|^2 \id x.
$$

We are using units in which Planck's constant $\hbar=1$ and the
fundamental unit of charge $e=1$. In these units the physical value of
the speed of light $c$ is approximately $137$ or more precisely the
reciprocal of the fine structure constant $\alpha$.

Many objections may of course be raised about the Hamiltonian $H_N$.
Even with the relativistic kinetic energy operator the Hamiltonian is
not truly relativistically invariant.  For spin 1/2 fermions the
appropriate kinetic energy operator would rather be the Dirac
operator.  Also the magnetic field is treated classically as opposed to as
a quantized field. Unfortunately, we do not today know how to describe
a mathematically consistent fully relativistic model in which to even
formulate the problem of stability. As is well known the Dirac
operator is not bounded from below and introducing quantized fields
requires a renormalization scheme. These questions are being actively
studied at present, but we shall, as already mentioned, not discuss
them here (see \cite{BuFeFrGr,FeFrGr,LiLo2}.  One approach to the Dirac
operator would be to restrict to positive energy solutions for the
free Dirac operator.  For the non-magnetic case this would lead to a
problem essentially identical to the problem with the relativistic
kinetic energy $T^{\rm Rel}$.  In the magnetic field case this is
somewhat more difficult (see \cite{LiSiSo}).

We now turn to the important question of which Hilbert Space the
operator $H_N$ acts on.  The operator will be unbounded, but
we shall always consider the Friedrichs' extension of the restriction
to $C^\infty$ functions with compact support. Since the problem of
stability is a question of lower bounds, the Friedrichs' extension is
the correct setting.

We begin by discussing the Hilbert spaces for just one particle, {\it
  the one-particle space}. We consider, in general the one particle
spaces to be the square integrable functions corresponding to
particles with $q$ internal states, i.e., $\cH^q_1=L^2(\R^3;\C^q)$,
where the non-negative integer $q$ may be different for the different
particles. In terms of spin, this corresponds to particles
of spin $(q-1)/2$.

The many particle space of interest is then
\begin{equation}\label{eq:hilbertspaceph}
  \cH_N^{\rm Physical}=\left(\bigwedge^{N-K}\cH^q_1\right)
  \otimes\left(\bigotimes_{j=N-K+1}^N\cH^{q_j}_1\right),
\end{equation}
where $K$ is a positive integer.  In order for the Hamiltonian $H_N$
to act on the above Hilbert space we must require that all the masses
and charges of the particles $j=1,\ldots,N-K$ are the same. Put
differently, these particles are identical fermions. For simplicity we
assume that $m_j=1$ and $z_j=-1$, for all $j=1, \ldots, N-K$.  (Except
for the sign of charge this simply amounts to a choice of units).
Thus the fermions are negatively charged. We  assume the
remaining particles to be positively charged, i.e., $z_j>0$ for
$j=N-K+1,\ldots,N$. Moreover, all the fermions must be described by
the same kinetic energy operator. The positively charged particles
could, in principle, have any kinetic energy operators and even simply
vanishing kinetic energy (equivalent to infinite mass).  Of course,
the Pauli kinetic energy operator requires the spin of the particle to
be $1/2$, i.e., $q=2$.

Since fermions have half-integer spin one would maybe like to restrict
the negatively charged fermions accordingly, i.e., the corresponding
$q$ to be even. {F}or the discussion here this restriction, however,
plays no role.

The reader may ask why one does not consider the situation when all or
some of the positively charged particles are fermions or bosons. In
fact, stability in this case is a simple consequence of the situation
discussed above, since considering fermions or bosons simply means
restricting to certain subspaces.

To study what happens when we ignore Fermi statistics all together we
shall consider all particles to be described by the standard
non-relativistic kinetic energy $T$ and all to have mass $m_j=1$.
Moreover, we assume that all the charges $z_j$ are either $+1$ or
$-1$. We indeed consider the charge of each particle to be a
variable.  Put differently, we consider the $N$-particle Hilbert
space
\begin{equation}\label{eq:hilbertspace}
  \cH_N=\bigotimes_{j=1}^NL^2(\R^3\times\{1,-1\}),
\end{equation}
where the set $\{1,-1\}$ refers to the charge variable. 

We define the ground state energy as the bottom of the spectrum of the
Hamiltonian. We will not here address the question of whether this is
and eigenvalue or not, i.e., whether there really is a ground state.
When a magnetic field is present we consider the vector potential 
as a dynamic variable and we also minimize over it.
Thus
\begin{equation}\label{eq:energyph}
  E_N^{\rm Physical}=\inf_{\boldsymbol A}\inf\hbox{spec}_{\cH^{\rm physical}_N}H_N
  =\inf_{\boldsymbol A}\inf\biggl\{\frac{(\Psi,H_N\Psi)}{(\Psi,\Psi)}\ \biggl|\  
  \Psi\in C_0^\infty\cap\cH^{\rm physical}_N\setminus\{0\}\biggr\}
\end{equation}
and
\begin{equation}\label{eq:energy}
  E_N=\inf\hbox{spec}_{\cH_N}H_N=\inf\biggl\{\frac{(\Psi,H_N\Psi)}{(\Psi,\Psi)}\ \biggl|\  
  \Psi\in C_0^\infty\cap\cH_N\setminus\{0\}\biggr\}.
\end{equation}
Here $C_0^\infty$ refers to smooth functions of compact support. 
In (\ref{eq:energyph}) we minimize over all smooth vector potentials for
which $|\nabla\times{\boldsymbol A}|\in L^2(\R^3)$. Of course the minimization over 
the vector potential is only relevant if the kinetic energy really depends on 
${\boldsymbol A}$ otherwise the minimum simply occurs for ${\boldsymbol A}=0$.

The energy $E_N$ on the space where we consider the charges as
variables is precisely the same as we would get if we calculate the
energy for any fixed choice of charges and afterward minimize over
this choice. Note, in particular, that the charge variable commutes
with the Hamiltonian $H_N$. The reason for including the charges as
variables is that the Hamiltonian is then fully symmetric in all $N$
particles and not just in the positively or negatively charged
particles separately.

Strictly speaking the energies in (\ref{eq:energyph}) and
(\ref{eq:energy}) are only defined as the bottom of the spectrum if
the rightmost expressions are finite (i.e., not $-\infty$). It is in
this case that we can define the operators as Friedrichs' extensions.
The property that the ground state energies are bounded below is often
referred to as {\it stability of the first kind}. It holds except in
the cases when the fermions are described by the kinetic energy
operators $T^{\rm Rel}$ or $T^{\rm Pauli}$.  In these two cases
stability of the first kind requires that
$\max\{z_j\}/c=\max_j\{z_j\}\alpha$ is small enough in the case of
$T^{\rm Rel}$ or that $q\max_j\{z_j\}/c^2=q\max_j\{z_j\}\alpha^2$ is
small enough in the case of $T^{\rm Pauli}$.

\section{STABILITY AND INSTABILITY OF MATTER} 

Stability of matter is a stronger statement than stability of the first kind.
It means that there exists a constant $C\in\R$ such that 
\begin{equation}\label{eq:stabilityofmatter}
  E_N^{\rm Physical}\ge -CN
\end{equation}
for all $N$, i.e., that the total binding energy per particle is bounded. 
\begin{theorem}\label{thm:stabmat} {\bf\emph{(Stability of Matter)}}\hfill\\
  On the space $\cH_N^{\rm Physical}$ stability of matter
  (\ref{eq:stabilityofmatter}) holds with a constant $C$ that depends
  only on $q$ and $\max\{z_j\}$ if anyone of the following situations hold.
  \begin{itemize}
  \item The fermions are described by the
    standard non-relativistic kinetic energy $T$ or by the magnetic kinetic energy 
    $T^{\rm Mag}$.
  \item The fermions are described by the relativistic kinetic energy $T^{\rm Rel}$ 
    and $q\alpha(=qc^{-1})$  is small enough and $\max_j\{z_j\}\alpha\leq 2/\pi$.
  \item The fermions are described by the Magnetic Pauli kinetic energy $T^{\rm Pauli}$ 
    and $q\alpha(=qc^{-1})$  and $q\max_j\{z_j\}\alpha^2$ are small enough.
  \end{itemize}
\end{theorem}

The case of the standard non-relativistic kinetic energy is the
situation first settled by Dyson and Lenard~\cite{DyLe2} and later by
Lieb and Thirring~\cite{LiTh} with a constant of the correct order of
magnitude. There was also a proof by Federbush~\cite{Fed}.

Stability for the magnetic kinetic energy is an immediate consequence
of the non-magnetic case and the diamagnetic inequality. In fact, the
ground state energy is achieved without a magnetic field.

Stability for the relativistic kinetic energy was first solved by
Conlon~\cite{Co} and then improved by Fefferman and
de~La~Llave~\cite{FeLl}.  The version formulated here (which is sharp
with respect to the bound $2/\pi$) is due to Lieb and Yau~\cite{LiYa}.
The case of one electron and one nucleus had been studied previously by
Herbst~\cite{He} and Weder~\cite{We} and the case of one electron and
several nuclei by Lieb and Daubechies~\cite{DaLi}.

A proof of stability for the magnetic Pauli kinetic energy was first
published by Lieb, Loss, and Solovej~\cite{LiLoSo}, but had been
previously announced, (although with weaker bounds) by Fefferman (only
later published in \cite{Fe}). The result as formulated here is
optimal in the sense that if either $q\alpha$ or
$q\max_j\{z_j\}\alpha^2$ are large then stability does not hold.  This
was realized in a series of papers \cite{FrLiLo,LiLo,LoYa}.

In the case without Fermi statistics Dyson~\cite{Dy} proved that there
is no stability and he, in fact, conjectured the following result.
\begin{theorem}\label{thm:Dyson} {\bf\emph{(The asymptotic energy of a charged Bose gas)}}\hfill\\
For the energy $E_N$ defined in (\ref{eq:energy}) we have the asymptotics
\begin{equation}\label{eq:dyson}
  \lim_{N\to\infty}\frac{E_N}{N^{7/5}}=\inf\biggl\{{\textstyle\frac{1}{2}}
  \int|\nabla\phi|^2-J\int\phi^{5/2}\ \biggl|\ \phi\geq0,\ \int\phi^2=1\biggr\},
\end{equation}
where
$$
J
=\left(\frac{2}{\pi}\right)^{3/4}\int_0^\infty1+x^4-x^2\left(x^4+2\right)^{1/2}\,dx
=\left(\frac{4}{\pi}\right)^{3/4}\frac{\Gamma(\frac{1}{2})\Gamma(\frac{3}{4})}{5\Gamma(\frac{5}{4})}.
$$
\end{theorem}
The reader may wonder why the theorem refers to a charged {\it Bose}
gas when, in fact, no statistics was enforced in the definition of the
Hilbert space $\cH_N$.  The reason is that the ground state energy on
$\cH_N$ is the same as the ground state energy one would get if
restricted to the fully symmetric subspace (i.e., the bosonic
subspace).  This is a fairly simple consequence of the facts that the
Hamiltonian is fully symmetric in all variables and that the expected
energy of any trial state does not increase if we replace it by its
absolute value.

It is a simple consequence of the classical Sobolev inequality (see
below) that the right side of (\ref{eq:dyson}) is finite. Dyson proved
in \cite{Dy} that $E_N\leq -CN^{7/5}$, but not with the correct
constant. His method and conjecture was inspired by the Bogolubov
approximation, which had been previously used by Foldy~\cite{Fo} to
calculate the energy asymptotics for the high density one-component Bose
plasma (bosonic jellium).  The Bogolubov approximation is usually applied to
bosonic systems, it is therefore important that we can think of our
system as such.

In their original work on stability of matter \cite{DyLe1} Dyson and
Lenard obtained as a corollary a lower bound on the Bose energy $E_N
\geq -C'N^{5/3}$ (see also Brydges and Federbush~\cite{BrFe}). The
correct exponent $7/5$ was however first proved by Conlon, Lieb and
Yau in \cite{CoLiYa}, where one finds the lower bound $E_N\geq
-C''N^{7/5}$, but again not with the correct constant.

Dyson's conjecture was finally proved in two papers establishing
respectively an upper and a lower bound.  An asymptotic lower bound of
the form (\ref{eq:dyson}) was proved by Lieb and
Solovej~\cite{LiSoII}.  The corresponding asymptotic upper bound will
appear in Solovej~\cite{So}.

The rigorous calculation of Foldy's high density asymptotics for the
energy of the one-component plasma can be found in Lieb and
Solovej~\cite{LiSoI} (lower bound) and Solovej~\cite{So} (upper
bound).

The main goal here is to review the techniques used to prove Dyson's
conjecture and to sketch the main steps in the proof.  Many of the
techniques used were developed to study stability of matter and I will
therefore use the opportunity to briefly review these applications as
well.

The main strategy in proving the lower bounds in
Theorems~\ref{thm:stabmat} and \ref{thm:Dyson} is to estimate the
Hamiltonian below by a simplified Hamiltonian for which one can get a
fairly explicit lower bound.  For Theorem~\ref{thm:stabmat} the
simplified Hamiltonian is a non-interacting mean field type
Hamiltonian. For Theorem~\ref{thm:Dyson} the simplified Hamiltonian is
a non-particle conserving Hamiltonian, which in the language of second
quantization is quadratic in creation and annihilation operators (a
quadratic Hamiltonian).

We first discuss how to treat the simplified Hamiltonians.

\section{ONE-PARTICLE OPERATORS AND QUADRATIC HAMILTONIANS}
The most basic estimate on the ground state energy of a one-particle
Hamiltonian, given in the next theorem, is a simple consequence of the
Sobolev inequality $\int|\nabla\psi|^2\geq C(\int|\psi|^6)^{1/3}$,
$C>0$ for functions $\psi$ on $\R^3$.
\begin{theorem}\label{thm:Sobolev} {\bf\emph{(Sobolev estimate)}}\hfill\\
  Let $V$ be a locally integrable function on $\R^3$, such that the
  Schr\"odinger operator $-\Delta+V$ may be defined as a quadratic
  form on $C^\infty$ functions with compact support. Then
$$
-\frac{1}{2}\Delta+V\geq-C_{\rm S}\int|V(x)|_-^{5/2}dx
$$
where $C_{\rm S}>0$ and $|t|_-=\max\{-t,0\}$.
\end{theorem}
This theorem of course immediately implies that the many-body operator
$$\sum_{i=1}^N-\frac{1}{2}\Delta_i+V(x_i)$$ 
has the lower bound $-C_{\rm S}N\int|V(x)|_-^{5/2}dx$.
On the fermionic subspace we, however, have the much stronger Lieb-Thirring inequality \cite{LiTh}.  
\begin{theorem}\label{thm:LiebThirring} {\bf\emph{(Lieb-Thirring inequality)}}\hfill\\
  On the fermionic space $\bigwedge^{M}\cH_1^q$ ($M$ a positive integer) we
  have the operator inequality
  $$
  \sum_{i=1}^M -\frac{1}{2}\Delta_i+V(x_i)\geq -C_{\rm LT}q\int
  |V(x)|_-^{5/2}dx,
  $$
  where $C_{\rm LT}>0$ $V$ is a locally
  integrable function on $\R^3$.
\end{theorem}

For the relativistic operator $T^{\rm Rel}$ and the magnetic Pauli
operator $T^{\rm Pauli}$ one has similar inequalities on the space
$\bigwedge^{M}\cH_1^q$ ($q=2$ for $T^{\rm Pauli}$). For $T^{\rm Rel}$ (with $m=1$) Daubechies~\cite{Da} proved
\begin{equation}\label{eq:daubechies}
\sum_{i=1}^M T_i^{\rm Rel}+V(x_i)\geq -C_{\rm D}qc^{-3}\int|V(x)|_-^{4}dx-Mc^{2}.
\end{equation}
For $T^{\rm Pauli}$ (with $m=1$ and $z=1$) the inequality
\begin{eqnarray}\label{eq:pauliLT}
  \sum_{i=1}^M T_i^{\rm Pauli}+V(x_i)&\geq& -C^{(1)}_{\rm LLS}\int
  |V(x)|_-^{5/2}dx\nonumber\\&&  -C^{(2)}_{\rm LLS}c^{-3/2}\left(\int|\nabla\times{\boldsymbol A}|^2dx\right)^{3/4}
  \left(\int|V(x)|_-^{4}dx\right)^{1/4},
\end{eqnarray}
can be found in \cite{LiLoSo}.

When studying bosonic systems one may, as explained above, use the
Sobolev estimate to get a lower bound on the energy. This will in
general not give the best dependence on the number of particles.
E.g., for the charged gas this would lead to a lower bound
$-CN^{5/3}$ as in \cite{DyLe1,BrFe}.  To get the sharp dependence
$-CN^{7/5}$ a more precise treatment of the interplay between the kinetic
energy and the Coulomb potential is needed.

This brings us to Bogolubov's method for calculating the energy of a Bose
gas.  In the Bogolubov approximation one assumes that most particles
form a Bose condensate in a momentum zero state.  The main
contribution to the ground state energy will come from the correlation
between two non-condensed particles with opposite momenta.  This
effect is most easily explained using the formalism of creation and
annihilation operators.  The following theorem gives a rigorous
formulation of Bogolubov's method in the simple case used by Foldy in
\cite{Fo}.
\begin{theorem}\label{thm:FoldyBogolubov} {\bf\emph{(Bogolubov's method)}}\hfill\\
  Assume that $\bn_{\pm,\pm}$ are four (unbounded) operators defined
  on a common dense domain on a Hilbert space, such that their
  adjoints are also defined on the same domain. Assume moreover that
  in the sense of quadratic forms on this domain we have the commutator identities
  $$
  \left[b^*_{\tau',z'},b^*_{\tau,z}\right]=\left[\bn_{\tau',z'},\bn_{\tau,z}\right]
  =\left[\bn_{\tau',-},b^*_{\tau,+}\right]=\left[\bn_{\tau',+},b^*_{\tau,-}\right]=0,\quad\hbox{for
    all }z,z',\tau,\tau'=\pm,
  $$
  and
  $$
  \quad\left[\bn_{\tau,z},b^*_{\tau,z}\right]\leq1.  \quad\hbox{for
    all }z,\tau=\pm.
  $$
  Then for all $t,g_+,g_-\geq0$ we have (again in the sense of quadratic forms) 
  \begin{eqnarray*}
    t\sum_{\tau,z=\pm1}b_{\tau,z}^*\bn_{\tau,z}+
    \sum_{z,z'=\pm1}\sqrt{g_zg_{z'}}zz'(b^*_{+,z}\bn_{+,z'}+b^*_{-,z}\bn_{-,z'}+b^*_{+,z}b^*_{-,z'}
    +\bn_{+,z}\bn_{-,z'})\\ \geq
    -(t+g_++g_-)+\sqrt{(t+g_++g_-)^2-(g_++g_-)^2}.
  \end{eqnarray*}
\end{theorem}
The operator inequality in this theorem follows simply by completing
squares (see also~\cite{LiSoII}). The importance of the lower bound is
emphasized by the fact that it is sharp if the operators $b_{\pm,\pm}$
are truly annihilation operators. In Bogolubov's method the operators
are however not exactly annihilation operators. Rather one should
think of $b_{\pm,z}$ as the operator that annihilates a particle with
charge $z$ and momentum $\pm$ (here only the sign of the momentum is
important) and creates a particle in the condensate (i.e., with
momentum 0) with charge $z$. The value $t$ represents the kinetic
energy of the particles with momentum $\pm$ and $g_\pm$ represent the
strength of the Coulomb interaction.

\section{REDUCTION TO A SIMPLIFIED HAMILTONIAN}\label{sec:reduction}

The reduction to a simplified Hamiltonian require controlling the
Coulomb interaction. It is often convenient to replace the Coulomb
potential by the Yukawa potential
$$
Y_\mu(x)=\frac{\exp(-\mu|x|)}{|x|}, \ \mu\geq0.
$$
Replacing the Coulomb potential $Y_0$ by a Yukawa potential $Y_\mu$
with $\mu>0$ amounts to introducing a long distance cut-off in the
potential. It is easy to control this replacement since $Y_0-Y_\mu$
has positive Fourier transform (is of positive type). Hence
$$
\sum_{1\leq i<j\leq
  N}z_iz_j(Y_0-Y_\mu)(x_i-x_j)\geq-\sum_{i=1}^N\frac{z_i^2\mu}{2},
$$
since $(Y_0-Y_\mu)(0)=\mu$. 

In order to bound the many-body Hamiltonian $H_N$ below by a one-body
Hamiltonian one must estimate the two-body potential by a one-body
potential.
\begin{theorem}\label{thm:Onsager} {\bf\emph{(Onsager's correlation estimate)}}\hfill\\
  Given particles with with charges $z_1,\ldots,z_N$ at positions
  $x_1,\ldots,x_N$. Let $$
  D_i=\min\{|x_i-x_j|\ |\ z_iz_j<0\},
  $$
  i.e. $D_i$ is the shortest distance from the particle at $x_i$ to a
  particle with the opposite charge. Then
  $$
  \sum_{1\leq i<j\leq N}z_iz_jY_\mu(x_i-x_j)\geq -\sum_{i=1}^N
  z_i^2\left(\frac{1}{12}(D_i\mu)^2+\frac{1}{2}(D_i\mu)+1\right)Y_\mu(D_i).
  $$
\end{theorem}
Above $D_i$ depends only on $x_i$ and on the positions of all the
particles with the opposite charge of the particle at $x_i$.  This
theorem allows us to separate the original Hamiltonian $H_N$ in a
one-body Hamiltonian for all the negatively charged particles (with a
one-body potential depending on the positions of all the positively
charged particles) and a one-body Hamiltonian for all the positively
charged particles (with a one-body potential depending on the
positions of all the negatively charged particles).
 
The proof of this theorem essentially goes back to Onsager~\cite{On}
(for $\mu=0$), who addressed a stability question for classical
charged matter.  In \cite{LiSoII} it is being used to introduce a
short distance cutoff in the potential (i.e., to replace $Y_0$ by
$Y_0-Y_\mu$ for large $\mu$).  The one-body potential is controlled
using the Sobolev estimate Theorem~\ref{thm:Sobolev}.

If both the negatively and positively charged particles are fermions,
we may use Onsager's estimate together with the Lieb-Thirring
inequality in Theorem~\ref{thm:LiebThirring} to prove stability of
matter.  This special case was treated by Dyson and Lenard in their
first paper \cite{DyLe1}, where they also used a version of Onsager's
correlation estimate, but did not have the Lieb-Thirring
inequality at their disposal.

In order to prove stability of matter as formulated in
Theorem~\ref{thm:stabmat} one may use the following stronger
correlation estimate due to Baxter~\cite{Ba}.
\begin{theorem}\label{thm:LY} {\bf\emph{(Baxter's correlation estimate)}}\hfill\\
Assume as before that all negatively charged particles have charge $-1$ Then 
$$
\sum_{1\leq i<j\leq N}\frac{z_iz_j}{|x_i-x_j|}\geq -\sum_{i=1, z_i<0}^N (1+2\max\{z_i\})D_i^{-1}.
$$
\end{theorem}
The importance here is that the sum on the right is only over
negatively charged particles.  Thus the right side is a one-body
potential for the negatively charged particles thinking of the
positively charged particles as fixed.  A sharper (although more
complicated to state) estimate was given by Lieb and Yau in
\cite{LiYa}. The estimate of Lieb and Yau is sharper in the sense that
the coefficient $1+2z$ above may be changed to $z$, but additional
bounded errors are needed.

\begin{remark}{\bf{(The proof of stability of matter)}}\hfill\\
A proof of stability of matter in the standard non-relativistic case 
may proceed as follows. We use Baxter's estimate to arrive at
\begin{eqnarray*}
  H_N&\geq&\sum_{i=1, z_i<0}^N(T_i-(1+2\max\{z_i\})(D_i^{-1}-R^{-1}))-
  N(1+2\max\{z_i\})R^{-1} \\&\geq&-C N(1+2\max\{z_i\})^{2},
\end{eqnarray*}
where we have used the Lieb-Thirring inequality
Theorem~\ref{thm:LiebThirring} for the potential
$$V=(1+2\max\{z_i\})(-D_i^{-1}+R^{-1})$$ and we chose
$R\sim(1+2\max\{z_i\})^{-1}$. 

The proof of stability of matter by Lieb and Thirring\cite{LiTh} used
the No-binding Theorem of Thomas-Fermi theory (see Lieb and
Simon~\cite{LiSi}) instead of Baxter's estimate.

If the fermions are described by the relativistic kinetic energy
$T^{\rm Rel}$ or the magnetic Pauli operator $T^{\rm Pauli}$ the proof
of stability is not quite as simple.  Note in particular that a
potential that has a Coulomb singularity is not integrable to the
fourth power and one can therefore not directly use the inequalities
(\ref{eq:daubechies}) or (\ref{eq:pauliLT}). The reader is referred to
\cite{LiYa} and \cite{LiLoSo} for the detailed proofs of Stability in
these cases.
\end{remark}

\subsection{MANY-BODY LOCALIZATION TECHNIQUES}
In many aspects of many-body theory the method of localizing has been
a very useful technique. Dyson and Lenard used it heavily in their
papers \cite{DyLe1,DyLe2}.

We here discuss the two problems connected with localizing a charged system, i.e., 
the localization of the interaction and the localization of the kinetic energy.

The simplest way to approach the localization of the kinetic energy is
to do a Neumann localization.  In fact, let $-\Delta_{\rm Neu}^{(y)}$
be the Neumann Laplacian for a unit cube centered at $y\in\R^3$.  Then in
terms of quadratic forms defined on smooth compactly supported
functions we may write
\begin{equation}\label{eq:Neumann}
  -\Delta=\int_{\R^3}-\Delta_{\rm Neu}^{(y)} dy ,
\end{equation}
where $-\Delta$ is the Laplacian on all of $\R^3$. Of course, there is
a rescaled version to cubes of other sizes. The structure of this
identity, i.e., that we have written the original operator as an
integral over localized operators is characteristic for the way we
shall write localizations in this section. The idea is then that for
each value of the integration parameter one has a localized problem
that one will estimate. Afterward the estimate is integrated.

Another approach is to use a smooth localization.  Let $\upchi$ be a
$C^\infty$ function with compact support. Let
$\upchi_y(x)=\upchi(x-y)$.  Then we have the identity
\begin{equation}\label{eq:IMS}
  \left(\int\upchi^2\right)(-\Delta)=\int_{\R^3} -\upchi_y\Delta\upchi_y dy-\int\nabla\upchi^2,
\end{equation}
which is often referred to as the IMS localization formula (see
\cite{CyFrKiSi}). Instead of writing the localization as an integral
one often uses a sum instead, but then one must introduce a partition
of unity.

For the problem of the charged Bose gas it turns out that we cannot
use either of these localization methods.  In this case we cannot
completely localize the kinetic energy operator. We must still have
some kinetic energy to control the variation between localized
regions.  The solution is to write the kinetic energy as a sum of a a
high energy part and a low energy part. We then localize only the high
energy part and use the low energy part to control the large distance
variation. The result can be found in \cite{LiSoII}.  We shall state a
simplified version here.

Assume that $\upchi$ with $0\leq\upchi\leq1$ is smooth has support in
a unit cube centered at the origin in $\R^3$ and that
$\sqrt{1-\upchi^2}$ is also smooth. Let as before
$\upchi_y(x)=\upchi(x-y)$.  Let $a^*(y)$ be the creation operator for
a particle in a normalized constant function in the unit cube centered
at $y\in\R^3$. Let $\cP_y$ be the orthogonal projection operator
projecting on functions orthogonal to constants in the cube.  Then for
any bounded set $\Omega\subset\R^3$ and any $0<s<1$ we have the
following many-body kinetic energy localization estimate on the
symmetric tensor product space $\bigotimes^N_{\rm sym} L^2(\R^3)$ (the
fully symmetric subspace of the full tensor product space)
\begin{eqnarray}
  (1+\varepsilon(\upchi,s))\sum_{i=1}^N
    -\Delta_i
  &\geq& \int_{\Omega}\Bigl[\sum_{i=1}^N\cP_y^{(i)}\upchi_{y}^{(i)}
    \frac{(-\Delta_i)^2}{-\Delta_i
      +s^{-2}}\upchi^{(i)}_y\cP^{(i)}_y
    \nonumber\\&&
    +\sum_{j=1}^3\left(\sqrt{a^*_0(y+e_j)\an_0(y+e_j)+1/2}-
      \sqrt{a^*_0(y)\an_0(y)+1/2}\right)^2
  \Bigr]dy\nonumber\\&&-3\hbox{\rm vol}(\Omega), \label{eq:manybodykin}
\end{eqnarray}
where $e_1,e_2, e_3$ is the standard basis in $\R^3$ and
$\varepsilon(\upchi,s)\to0 $ as $s\to0$.  Note that this is an
explicit many-body bound in the sense that the right side of the
operator inequality is not a one-body operator.  The above estimate
generalizes immediately to the situation where a charge variable is
present.  In this case there should be two creation operators, one for
charge $+1$ and one for charge $-1$.  There are rescaled version of
(\ref{eq:manybodykin}), which hold if we replace the unit cube by
cubes of other sizes.

The first sum on the right side of (\ref{eq:manybodykin}) is the
localization of the high energy part of the kinetic energy. High
energy means much larger than $s^{-2}$. 

The second sum on the right side of (\ref{eq:manybodykin}) represents
the kinetic energy due to the variations between localized regions.
Using the Cauchy-Schwarz inequality one sees that for a state
$\Psi\in\bigotimes^N_{\rm sym} L^2(\R^3)$ we have
\begin{eqnarray*}
 \lefteqn{\left(\Psi,\left(\sqrt{a^*_0(y+e_j)\an_0(y+e_j)+1/2}-
      \sqrt{a^*_0(y)\an_0(y)+1/2}\right)^2\Psi\right)}&&\\
  &&\geq\left(\sqrt{\left(\Psi,(a^*_0(y+e_j)\an_0(y+e_j)+1/2)\Psi\right)}-
      \sqrt{\left(\Psi,(a^*_0(y)\an_0(y)+1/2)\Psi\right)}\right)^2.
\end{eqnarray*}
This allows us to think of the last term on the right side of
(\ref{eq:manybodykin}) as a discrete Laplacian acting on the function
\begin{equation}\label{eq:sqrt}
  y\mapsto\sqrt{\left(\Psi,(a^*_0(y)\an_0(y)+1/2)\Psi\right)}.
\end{equation}

For the Bose gas we shall conclude that most particles are in the
constant function state (the zero momentum state). More precisely,
this means that $(\Psi,a^*_0(y)\an_0(y)\Psi)$ is almost the expected
total number of particles in the unit cube centered at $y$. The
expectation of the last term on the right side of
(\ref{eq:manybodykin}) will therefore essentially give a contribution
equal to the Laplacian of the square root of the density (assuming
that we can approximate the discrete Laplacian by the continuous
Laplacian).

We finally come to the discussion of the localization of the
interaction.  For the Coulomb or Yukawa interaction this can be done
using a method of Conlon, Lieb, and Yau~\cite{CoLiYa}.  Let as before
$\upchi$ be a smooth function supported in the unit cube centered at
the origin. There is an $\omega>0$ depending on $\upchi$ such that for
all $\mu\geq0$ we have
\begin{equation}\label{eq:CLY}
  \left(\int\upchi^2\right)\sum_{1\leq i<j\leq N}z_iz_j
  Y_\mu(x_i-x_j) \geq\int_{\R^3}\sum_{1\leq i<j\leq
    N}\upchi_y(x_i)Y_{\mu+\omega}(x_i-x_j)\upchi_y(x_j)dy -N\omega.
\end{equation}
Note that the effect of the localization function $\upchi_y$ on the
right side is that for each fixed value of the integration parameter
$y$ only particles that live in the unit cube centered at $y$ interact
with each other. Again it is easy to see how this estimate changes
under rescaling.  A very elegant version of this estimate was given by
Graf and Schencker~\cite{GrSch}.
 
\section{THE LOWER BOUND IN DYSON'S CONJECTURE}

We shall now describe the main steps in the proof of the lower bound
in Theorem~\ref{thm:Dyson}. The reader should look in \cite{LiSoII}
for details.  In each step we shall ignore certain errors and we shall
not explain in details how these errors are estimated. In the detailed
proof the errors are, in fact, only estimated at the very end when the
main contributions have been identified.

It is first of all important to understand that there are two relevant
length scales in the problem.  A long scale $L_0$, which is the
diameter of the Bose gas and a short scale $\ell_0$ which is the
distance on which the Bogolubov pairs interact.  It will turn out that
$L_0\sim N^{-1/5}$ and $\ell_0\sim N^{-2/5}$.

{\bf Step. 1:} We first localize the whole system into a large cube of
size $L\gg L_0\sim N^{-1/5}$. On the other hand $L$ should not be too
large in order to allow us to control the volume error in
(\ref{eq:manybodykin}).  This first localization is done using the
Conlon-Lieb-Yau estimate (\ref{eq:CLY}) and an IMS-localization
(\ref{eq:IMS}) of the kinetic energy. Note that after rescaling the
error in (\ref{eq:CLY}) will be $N\omega/L\ll \omega N^{6/5}$.  The
IMS localization error (i.e., $N\int(\nabla\upchi)^2/\int\upchi^2$)
will be $N L^{-2}\ll N^{-7/5}$.
 
{\bf Step. 2:} We do a second localization into boxes of size $\ell$,
where $\ell_0\ll\ell\ll L_0$. This localization is done using the
many-body kinetic energy localization (\ref{eq:manybodykin}) together
with the Conlon-Lieb-Yau estimate (\ref{eq:CLY}).  This time the error
in using (\ref{eq:CLY}) is $N\omega/\ell\ll N\omega /\ell_0 \sim
\omega N^{7/5}$.

The result of this localization is that the total
ground state energy is estimated below by a sum of two terms.  One
term is the discrete Laplacian term discussed in the previous section.
The other term is the local energy, which will eventually lead to the
effective Hamiltonian for which we may apply the Bogolubov
approximation.

{\bf Step 3:} Before proceeding with the analysis of the local short
scale energy we introduce long and short distance cut-offs in the
interaction. This is done as explained in the beginning of
Section~\ref{sec:reduction}. 

{\bf Step 4:} This is the final step in the reduction to the effective
Hamiltonian. The localized two body potential is of the form
$W(x,x')=\upchi_y(x)V(x-x')\upchi_y(x')$, where $V$ is a cut-off Coulomb
potential, i.e, $V=Y_\mu-Y_\nu$ for some appropriate $\mu\ll\nu$. Here 
$\upchi_y$ has support in a cube of side $\ell$ centered at $y$. 

Let $a^*_{\alpha z}$ be the creation operator for a particle of charge
$z=\pm1$ in a state $u_\alpha$, $\alpha=0,1,\ldots$, where
$\{u_\alpha\}$ is an orthonormal basis for the Hilbert space of $L^2$
on the cube of size $\ell$ centered at $y$ and $u_0$ is the constant
function.  It is not otherwise important how this basis is chosen. We
may write the full two body interaction as
$$
\mfr{1}{2}\sum_{z,z'=\pm1}\sum_{\alpha\beta\gamma\delta}
zz'(u_\alpha\otimes u_\beta, W u_\beta\otimes u_\delta) 
a^*_{\alpha z}a^*_{\beta z'}\an_{\delta z'}\an_{\gamma z}.
$$
The fourth step in the proof is to show that one may ignore (for a
lower bound) all terms in this sum which do not have precisely two of
the parameters $\alpha,\beta,\gamma,\delta$ equal to zero. Moreover
one may also ignore terms for which $\alpha=\gamma=0$ or
$\beta=\delta=0$. This step in the proof is rather technical and uses
heavily that we have been able to introduce cut-offs in the potential.

The resulting two body interaction may be written 
$$
\frac{1}{2(2\pi)^3}\int_{p\in\R^3}\hat V(p)\sum_{z,z'=\pm1}\sqrt{\nu^z\nu^{z'}}
zz'(b^*_{+p,z}\bn_{+p,z'}+b^*_{-p,z}\bn_{-p,z'}+b^*_{+p,z}b^*_{-p,z'}
  +\bn_{+p,z}\bn_{-p,z'})\,dp,
$$
where $\hat V$ denotes the Fourier transform of the potential $V$,
$\nu^{\pm}$ denote the total number of positively and negatively
charged particles in the cube respectively, and 
$$
b^*_{p,z}=(\ell^3\nu^z)^{-1/2}a^*_z(\cP_y\upchi_y e^{ipx})a_{0z},
$$
where $a^*_z(f)$ creates a particle of charge $z$ in the state $f$
and as before $\cP_y$ is the projection orthogonal to constants. It
is easy to see that for fixed $p$ the operators $b^*_{\pm p,z}$
satisfy the conditions in Theorem~\ref{thm:FoldyBogolubov} on the 
Foldy-Bogolubov method.

It is also easy to see that the localized kinetic energy resulting from 
(\ref{eq:manybodykin}) may be estimated below by 
$$
\frac{1}{2(2\pi)^{3}}\int_{p\in\R^3}t(p)\sum_{\tau,z=\pm1}b_{\tau p,z}^*\bn_{\tau p,z}\, dp,
$$
where $t(p)=\frac{1}{2}\ell^3 p^4(p^2+\ell s^{-2})^{-1}$.

It is now immediate from the Foldy-Bogolubov method in
Theorem~\ref{thm:FoldyBogolubov} that the local ground state energy is
estimated below by
$$
-\frac{1}{2(2\pi)^{3}}\int_{\R^3}(t(p)+\nu_y \hat V(p))-\sqrt{(t(p)^2+2t(p)\nu_y \hat V(p)}\, dp
$$
where $\nu_y=\nu^++\nu^-$ (which depends on the location of the
cube, although this had been suppressed for $\nu^{\pm}$).  If we now
ignore the various cut-offs (which may be justified) and simply
replace $t(p)$ by $\frac{1}{2}\ell^3 p^2$ and $\hat V(p)$ by
$4\pi/p^2$ we see that the above integral  gives
$-J\nu_y^{5/4}\ell^{-3/4}=-J(\nu_y/\ell^3)^{5/4}\ell^{3}$, 
with $J$ defined in Theorem~\ref{thm:Dyson}.

{\bf Step 5:} In order to control the errors encountered in reducing
to the effective Hamiltonian as well as in treating the discrete
Laplacian term we must conclude that most particles are in the
condensate, as explained in the previous section.  To do this, we
shall localize the number of excited particles.  To be more precise,
we may think of the Hilbert space as being a direct sum over subspaces
with a definite number of particles in the condensate. The Hamiltonian
is not diagonal in this representation.  We want to conclude however
that restricting any given state to only a small finite number of the
subspaces will not significantly change its expected energy. This is
achieved using the following result from \cite{LiSoI}.

\begin{theorem}{\bf\emph{(Localization of large matrices)}}\hfill\\ \label{thm:local} 
  Suppose that ${\cA}$ is an $N\times N$ Hermitian matrix and let
  ${\cA}^k$, with $k=0,1,\ldots,N-1$, denote the matrix consisting of
  the $k^{\rm th}$ supra- and infra-diagonal of ${\cA}$.  Let $\psi
  \in \C^N$ be a normalized vector and set $d_k = (\psi , {\cA}^k
  \psi) $ and $\lambda = (\psi , {\cA} \psi) = \sum_{k=0}^{N-1} d_k$.
  \ ($\psi$ need not be an eigenvector of ${\cA}$.) \ 
  
  Choose some positive integer $M \leq N$.  Then, with $M$ fixed,
  there is some $n \in [0, N-M]$ and some normalized vector $ \phi \in
  {\C}^N$ with the property that $\phi_j =0$ unless $n+1 \leq j
  \leq n+M$ \ (i.e., $\phi $ has length $M$) and such that
  \begin{equation}\label{localerror}
    (\phi , {\cA} \phi) \leq \lambda + \frac{C}{ M^2}
    \sum_{k=1}^{M-1} k^2 |d_k|
    +C\sum_{k=M}^{N-1} |d_k|\ ,
  \end{equation}
  where $C>0 $ is a  universal constant. (Note that the first sum starts
  with $k=1$.)
\end{theorem}

{\bf Step 6:} The final step is to combine the two parts of the energy
described in Step.~2.  The local energy was (approximately) bounded below by
$-J(\nu_y/\ell^3)^{5/4}\ell^{3}$ which when integrated over $y$ and
normalized by the volume of the cube gives
$$
-J\int_{\R^3}\left(\nu_y/\ell^3\right)^{5/4}dy.
$$
The other part of the energy is essentially the kinetic energy of
the function in (\ref{eq:sqrt}) (where we had actually ignored the
charged variable). Using the result of Step~5 that most particles are
in the condensate we may write this kinetic energy as (approximately)
$$
\frac{1}{2}\int_{\R^3}\left(\nabla\sqrt{\nu_y/\ell^3}\right)^2dy.
$$
If we use that the total number of particles is $N$ we have the
condition that $\int_{\R^3}{\nu_y/\ell^3}\,dy=N$.  Thus if we define
$\phi(y)=N^{-4/5}\sqrt{\nu_{N^{-1/5}y}/\ell^3}$ we see by a
straightforward scaling argument that $\int\phi(x)^2dx=1$ and that the
ground state energy is approximately bounded below by
$$
N^{7/5}\left(\frac{1}{2}\int_{\R^3}\left(\nabla\phi(x)\right)^2dx-
J\int_{\R^3}\left(\phi(x)\right)^{5/2}dy\right),
$$
which is of course bounded below by minimizing over all $\phi$ as
in (\ref{eq:dyson}).
 
\section{THE UPPER BOUND IN DYSON'S CONJECTURE}

To prove an upper bound on $E_N$ of the form given in Dyson's
conjecture Theorem~\ref{thm:Dyson} we shall construct a trial function
using as an input a minimizer $\phi$ for the variational problem on
the right side of (\ref{eq:dyson}). That minimizers exist can be seen
using spherical decreasing rearrangements.  Define
$\phi_0(x)=N^{3/10}\phi(N^{1/5}x)$.

Let $\phi_\alpha$, $\alpha=1,\ldots$ be an orthonormal family of real
functions all orthogonal to $\phi_0$.  We choose these functions
below.

We follow Dyson~\cite{Dy} and first choose a trial function which does
not have a specified particle number, i.e., a state in the space
$\bigoplus_{N=0}^\infty\cH_N$, ($\cH_0=\C$) more precisely in the
bosonic subspace, i.e., the bosonic Fock space.  We shall evaluate the
expected value of the operator $\sum_{N=0}^\infty H_N$ in our state.

As our trial many-body wave function we now choose
\begin{equation}\label{eq:trialstate} 
\Psi=\prod_{\alpha\ne0}(1-\lambda_\alpha^2)^{1/4}\exp\left(-\lambda_0^2+\lambda_0a^*_{0+}+\lambda_0
a^*_{0-}-\sum_{z,z'=\pm1}\sum_{\alpha\ne
  0}\frac{\lambda_{\alpha}}{4} zz'a_{\alpha,z}^*a_{\alpha,z'}^*\right)\left|0\right\rangle,
\end{equation}
where $a_{\alpha,z}^*$ is the creation of a particle of charge
$z=\pm1$ in the state $\phi_\alpha$, $|0\rangle$ is the vacuum state,
and the coefficients $\lambda_0,\lambda_1,\ldots$ will be chosen
below satisfying $0<\lambda_\alpha<1$ for $\alpha\ne0$.

It is straightforward to check that $\Psi$ is a normalized function.

Dyson used a very similar trial state in \cite{Dy}, but in his case
the exponent was a purely quadratic expression in creation operators,
whereas the one used here is only quadratic in the creation operators
$a^*_{\alpha z}$, with $\alpha\ne0$ and linear in $a^*_{0\pm}$.  As a
consequence our state will be more sharply localized around the mean
of the particle number.

Consider the operator
$$
\Gamma=\sum_{\alpha=1}^\infty
\frac{\lambda_\alpha^2}{1-\lambda_\alpha^2}|\phi_\alpha\rangle\langle\phi_\alpha|.
$$ 
A straightforward calculation of the energy expectation in the state $\Psi$ gives that
\begin{equation}\label{eq:upper}
  \left(\Psi,\sum_{N=0}^\infty H_N \Psi\right)=\lambda_0^2 \int(\nabla \phi_0)^2+\hbox{Tr}\left(T\Gamma\right)
  +2\lambda_0^2\hbox{Tr}\left(\cK\left(\Gamma-\sqrt{\Gamma(\Gamma+1)}\right)\right),
\end{equation}
where $\cK$ is the operator with integral kernel
$$
\cK(x,y)=\phi_0(x)|x-y|^{-1}\phi_0(y).
$$
Moreover, the expected particle number in the state $\Psi$ is
$2\lambda_0^2+\hbox{Tr}(\Gamma)$. In order for $\Psi$ to be well
defined by the formula (\ref{eq:trialstate}) we must require this
expectation to be finite.

Instead of making explicit choices for the individual functions
$\phi_\alpha$ and the coefficients $\lambda_\alpha$, $\alpha\ne0$ we
may equivalently choose the operator $\Gamma$.  In defining $\Gamma$
we use the method of coherent states.  Let $\upchi$ be a non-negative
real and smooth function supported in the unit ball in $\R^3$, with
$\int\upchi^2=1$.  Let as before $N^{-2/5}\ll\ell\ll N^{-1/5}$ and
define $\upchi_\ell(x)=\ell^{-3/2}\upchi(x/\ell)$. We choose
$$
\Gamma(x,y)=(2\pi)^{-3}\int_{\R^3\times\R^3}f(u,|p|)\cP_{\phi_0}^\perp|\theta_{u,p}\rangle\langle\theta_{u,p}|\cP_{\phi_0}^\perp
dudp
$$
where $\cP_{\phi_0}^\perp$ is the projection orthogonal to
$\phi_0$,
$$
\theta_{u,p}(x)=\exp(ipx)\upchi_\ell(x-u),
$$
and
$$
f(u,|p|)=\frac{1}{2}\left(\frac{p^4+16\pi\lambda_0^2\phi_0(u)^2}{p^2\left(p^4+32\pi\lambda_0^2\phi_0(u)^2\right)^{1/2}}-1\right).
$$
We note that $\Gamma$ is a positive trace class operator, $\Gamma\phi_0=0$, and
that all eigenfunctions of $\Gamma$ may be chosen real. These are
precisely the requirements needed in order for $\Gamma$ to define the
orthonormal family $\phi_\alpha$ and the coefficients $\lambda_\alpha$
for $\alpha\ne0$.

We use the following version of the Berezin-Lieb inequality
\cite{Berezin72,Lieb73}.  Assume that $\xi(t)$ is an
operator concave function on $\R_+\cup\{0\}$ with $\xi(0)\geq0$.  Then if $Y$ is a
positive semi-definite operator we have
\begin{equation}\label{eq:berezinlieb}
  \hbox{Tr} \left(Y\xi(\Gamma)\right)\geq (2\pi)^{-3}\int
  \xi(f(u,|p|))\left(\theta_{u,p},\cP_{\phi_0}^\perp
    Y\cP_{\phi_0}^\perp\theta_{u,p}\right) dudp.
\end{equation}
We use this for the function $\xi(t)=\sqrt{t(t+1)}$.  If $Y=I$ then
(\ref{eq:berezinlieb}) holds for all concave function $\xi$ with
$\xi(0)\geq0$. Of course, if $\xi$ is the identity function then
(\ref{eq:berezinlieb}) is an identity.

This reduces proving an upper bound on the energy expectation
(\ref{eq:upper}) to the calculations of explicit integrals. After
estimating these integrals one arrives at the leading contribution (in
$N$)
\begin{eqnarray*}
  &&\lambda_0^2 \int(\nabla \phi_0)^2\\
  &&+\iint\left(\frac{1}{2}p^2+2\lambda_0^2\phi_0(u)^2\frac{4\pi}{p^2}\right)f(u,|p|)
  -\frac{4\pi}{p^2}2\lambda_0^2\phi_0(u)^2\sqrt{f(u,|p|)(f(u,|p|)+1)}\ dpdu\\
&=&\lambda_0^2 \int(\nabla \phi_0)^2-J\int (2\lambda_0^2)^{5/4}\phi_0^{5/2},
\end{eqnarray*}
where $J$ is as in (\ref{eq:dyson}). 

If we choose $\lambda_0=\sqrt{N/2}$ we get after a simple rescaling
that the energy above is $N^{7/5}$ times the right side of
(\ref{eq:dyson}) (recall that $\phi$ was chosen as the minimizer).
We also note that the expected number of particles is 
$$
2\lambda_0^2+\hbox{Tr}(\Gamma)=N+O(N^{3/5}),
$$
as $N\to\infty$.

The only remaining problem is to show how a similar energy could be
achieved with a wave function with a fixed number of particles $N$,
i.e., how to show that we really have an upper bound on $E_N$.  We
indicate this fairly simple argument here.

We construct a trial function $\Psi'$ as above, but with an expected
particle number $N'$ chosen appropriately close to but slightly
smaller than $N$.  Using that we have a good lower bound on the energy
$E_N$ for all $N$ we may, without changing the energy expectation
significantly, replace $\Psi'$ by a normalized wave function $\Psi$
that only has particle numbers less than $N$.  Since the function
$N\mapsto E_N$ is a decreasing function we see that the energy
expectation in the state $\Psi$ is, in fact, an upper bound to
$E_{N}$.

%%%%%%%%%%%%%%%%%%%%%%%%%%%%%%%%%%%%%%%%%%%%%%%%%%%%%%%%%%%%%
% Doing Acknowledgement                                               %
%%%%%%%%%%%%%%%%%%%%%%%%%%%%%%%%%%%%%%%%%%%%%%%%%%%%%%%%%%%%%%

%\section*{Acknowledgments}

%TYPE ACKNOWLEGDEMENTS HERE.


\begin{thebibliography}{00}

%%%%%%%%%%%%%%%%%%%%%%%%%%%%%%%%%%%%%%%%%%%%%%%%%%%%%%%%%%%%%
%                                                           %
% Command to used is:-                                      %
%                                                           %
%  \bibitem{REFERENCE_LABEL} AUTHORS NAMES,                 %
%  {\it JOURNAL'S NAMES}{\bf VOLUME NUMBER}, PAGE (YEAR).   %
%                                                           %
%  See example below.                                         %
%                                                           %
%%%%%%%%%%%%%%%%%%%%%%%%%%%%%%%%%%%%%%%%%%%%%%%%%%%%%%%%%%%%%
\bibitem{Ba}{Baxter, J. R.}, {Inequalities for potentials of particle
    systems}, {\it Illinois J.\  Math.} {\bf 24}, {645--652} (1980).

\bibitem{Berezin72}
Berezin F.~A.,
{\it Izv. Akad. Nauk, ser. mat.}, {\bf 36} (No. 5) (1972).
English translation: USSR Izv. {\bf 6} (No. 5) (1972)
and
Berezin F.~A.,
{General concept of quantization.}
{\it Commun. Math. Phys.} {\bf 40}, 153--174 (1975).
  
\bibitem{BrFe} {Brydges, David and Federbush, Paul}, {A note on energy
    bounds for boson matter}, {\it Jour.\  Math.\  Phys.} {\bf 17},
  2133--2134 (1976).
  
\bibitem{BuFeFrGr} Bugliaro, Luca, Fr\"ohlich, J{\"u}rg, and Graf, Gian
  Michele, Stability of quantum electrodynamics with nonrelativistic
  matter, {\it Phys.\ Rev.\ Lett.} {\bf 77}, 3494--3497 (1996).

\bibitem{Ch} Chandrasekhar, Subramanyan, {\it Phil.\ Mag.} {\bf 11}, 592 (1931).

\bibitem{Co} {Conlon, Joseph G.},
{The ground state energy of a classical gas},
{\it Commun.\ Math.\ Phys.} {\bf 94}, 439--458
(1984).
 
\bibitem{CoLiYa} {Conlon, Joseph G. and Lieb, Elliott H. and Yau, Horng-Tzer},
{The {$N\sp {7/5}$} law for charged bosons},
{Commun.\  Math.\ Phys.}
{\bf 116}, {417--448}
(1988).

\bibitem{CyFrKiSi}{Cycon, H. L. and Froese, R. G. and Kirsch, W. and Simon, B.},
     {Schr\"odinger operators with application to quantum mechanics
              and global geometry},
    {in \it Texts and Monographs in Physics},
   {Springer-Verlag}, {Berlin}, (1987).
   
\bibitem{Da} Daubechies, Ingrid, An uncertainty principle for fermions
  with generalized kinetic energy, {\it Commun.\  Math.\  Phys.} {\bf 90},
  511--520 (1983).

\bibitem{DaLi}{Daubechies, Ingrid and Lieb, Elliott H.}, {One-electron
    relativistic molecules with {C}oulomb interaction}, {Comm. Math.
    Phys.}  {\bf90}, 497--510 ({1983}). 
 
\bibitem{Dy} Dyson, Freeman J., Ground state energy of a finite system
  of charged particles, {\it Jour.\  Math.\  Phys.} {\bf 8}, 1538--1545
  (1967).

\bibitem{DyLe1} Dyson, Freeman J. and Lenard, Andrew, Stability of
  matter. I, {\it Jour.\  Math.\  Phys.} {\bf 8}, 423--434,
  (1967).

\bibitem{DyLe2} Dyson, Freeman J. and Lenard, Andrew, Stability of
  matter. II, {\it Jour.\  Math.\  Phys.} {\bf 9}, 698--711 (1968).

\bibitem{Fed} {Federbush, Paul},
{A new approach to the stability of matter problem. {I}, {II}},
{\it Jour.\  Math.\  Phys.}
{\bf 16}, 347--351
(1975); ibid.\  {\bf 16}  706--709 (1975).

\bibitem{Fe} Fefferman, Charles L.\ , Stability of matter with magnetic
  fields, {\it CRM Proc.\  Lecture Notes} {\bf 12}, 119--133 (1997)
  
\bibitem{FeFrGr} {Fefferman, Charles and Fr{\"o}hlich, J{\"u}rg and
    Graf, Gian Michele}, {Stability of ultraviolet-cutoff quantum
    electrodynamics with non-relativistic matter}, {\it Commun.\  Math.\ 
    Phys.} {\bf 190}, {309--330} {(1997)}.

\bibitem{FeLl} {Fefferman, Charles and de la Llave, Rafael}, {Relativistic
    stability of matter.\  {I}}, {\it Rev.\  Mat.\  Iberoamericana}, {\bf
    2}, {119--213} (1986).
 
\bibitem{FiRu} Fisher, Michael and Ruelle, David, The stability of
  many-particle systems, {\it Jour.\ Math.\ Phys.} {\bf 7}, 260--270
  (1966).

\bibitem{Fo} Foldy, Leslie L., Charged boson gas, {\it Phys.\ Rev.}
  {\bf 124}, 649--651 (1961); Errata {\it ibid} {\bf 125}, 2208 (1962).

\bibitem{FrLiLo} Fr{\"o}hlich, J{\"u}rg and Lieb, Elliott H.\  and Loss,
  Michael, Stability of {C}oulomb systems with magnetic fields.\  {I}.\ 
  {T}he one-electron atom, {\it Commun.\  Math.\  Phys.} {\bf 104},
  {251--270} ({1986}).
  
 
\bibitem{GrSch}Graf,  Gian Michele and Schenker, Daniel, On the Molecular Limit of
Coulomb Gases, {\it Commun.\ Math.\ Phys.} {\bf 174}, 215--227 (1995).
  
\bibitem{He} Herbst, Ira W.\ , Spectral theory of the operator
  {$(p\sp{2}+m\sp{2})\sp{1/2}-Ze\sp{2}/r$}, {\it Commun.\  Math.\  Phys.}
  {\bf 53}, {285--294} ({1977}).
 
\bibitem{Lieb73}
Lieb, Elliott H.,
The classical limit of quantum spin systems,
{\it Commun.\ Math.\ Phys.} {\bf 31}, 327--340 (1973).

\bibitem{LiLo} Lieb, Elliott H.\  and Loss, Michael, {Stability of
    {C}oulomb systems with magnetic fields.\  {II}.\   {T}he many-electron
    atom and the one-electron molecule}, {\it Commun.\  Math.\  Phys.} {\bf
    104}, {271--282} (1986).
 
\bibitem{LiLo2}{Lieb, Elliott H.\  and Loss, Michael}, {Stability of
    a model of relativistic quantum electrodynamics}, {\it Commun.\  Math.\ 
    Phys.} {\bf  228}, {561--588} (2002).

\bibitem{LiLoSo} Lieb, Elliott H.\  and Loss, Michael and Solovej, Jan
  Philip, Stability of matter in magnetic fields, {\it Phys.\  Rev.\ 
    Lett.} {\bf  75}, 985--989 (1995).

\bibitem{LiSiSo} Lieb, Elliott H.\  and Siedentop, Heinz and Solovej,
  Jan Philip, Stability and instability of relativistic electrons in
  classical electromagnetic fields, {\it Jour.\  Stat.\  Phys.} {\bf 89},
  37--59 (1997).  

\bibitem{LiSi}{Lieb, Elliott H.\  and Simon, Barry},
     {The {T}homas-{F}ermi theory of atoms, molecules and solids},
   {\it Adv.\  in Math.}
    {\bf 23},{22--116}
     (1977).
        
\bibitem{LiSoI} {Lieb, Elliott H.\  and Solovej, Jan Philip}, {G}round
  state energy of the one-component charged {B}ose gas, {\it {C}omm.\ 
  {M}ath.\  {P}hys.} {\bf 217}, 127--163 (2001); Erratum: {\it
  ibid} {\bf 225}, {219--221} (2002).
 
\bibitem{LiSoII} {Lieb, Elliott H.\  and Solovej, Jan Philip}, {G}round
  state energy of the two-component charged {B}ose gas, Preprint 2003.

\bibitem{LiTh} Lieb, Elliott H.\  and Thirring, Walter E.\ , Bound for the
  kinetic energy of fermions which proves the stability of matter,
  {\it Phys.\  Rev.\  Lett.} {\bf 35}, 687--689 (1975).
  
\bibitem{LiYa} Lieb, Elliott H.\  and Yau, Horng-Tzer, The stability and
  instability of relativistic matter, {\it Commun.\  Math.\  Phys.} {\bf
    118}, 177--213 (1988).
  
\bibitem{LoYa} Loss, Michael and Yau, Horng-Tzer, {Stabilty of
    {C}oulomb systems with magnetic fields.\  {III}.\   {Z}ero energy
    bound states of the {P}auli operator}, {\it Commun.\  Math.\  Phys.} 
  {\bf 104}, {283--290} (1986).

\bibitem{So} Solovej,  Jan Philip, In preparation.
   
\bibitem{On} Onsager, Lars, Electrostatic Interaction of Molecules,
  {\it Jour.\ Phys.\ Chem.} {\bf 43}, 189--196 (1939).

\bibitem{We} {Weder, R.\ A.}, {Spectral analysis of pseudodifferential operators},
 {\it J. Functional Analysis}
   {\bf 20} 319--337
  (1975).
  
\end{thebibliography}
\end{document}